 \documentclass[final,5p,times,twocolumn]{elsarticle}

\usepackage{graphicx}

\usepackage{amssymb}

\begin{document}

\begin{frontmatter}
\title{Si$\Lambda$ViO: A Trigger for $\Lambda$-Particles}
\author[L1]{Robert Muenzer\corref{cor1}}
\author[L1]{Laura Fabbietti}
\author[L1]{Martin Berger}
\author[L2]{Olaf Hartmann}
\address[L1]{Excellence Cluster Universe, Technische Universität¤t M\"unchen, D-85748, Germany}
\address[L2]{Stefan-Meyer-Institute for Subatomic Physics, Austrian Academy of Science, Vienna, Austria}
\cortext[cor1]{Present address: Boltzmannstr. 2, D-85748 Garching, Germany, Email address: rmuenzer@ph.tum.de}



\begin{abstract}
The production of bound kaon-nucleon-states (kaonic-cluster) in the reaction $p+p\rightarrow [ppK^{-}] + K^{+}$ should be investigated in a dedicated measurement at the FOPI detector system at the SIS-accelerator of GSI, by looking for the decay $ppK^{-} \rightarrow p + \Lambda \rightarrow p + p + \pi^{-}$.
To select the final states, the detector system Si$\Lambda$ViO has been build up as a trigger for $\Lambda$-Hyperons. This detector consists of two layers of segmented silicon-strip-detectors, which are able to measure online the particle multiplicity. These two layers are arranged such, that most of the $\Lambda$-Hyperons will decay between them. The trigger signal is generated by comparing online the hit multiplicity on the two layers.
\end{abstract}


\end{frontmatter}


\section{Introduction}
\label{Intro}
\begin{figure}[h]
\begin{center}
\includegraphics[width=5cm]{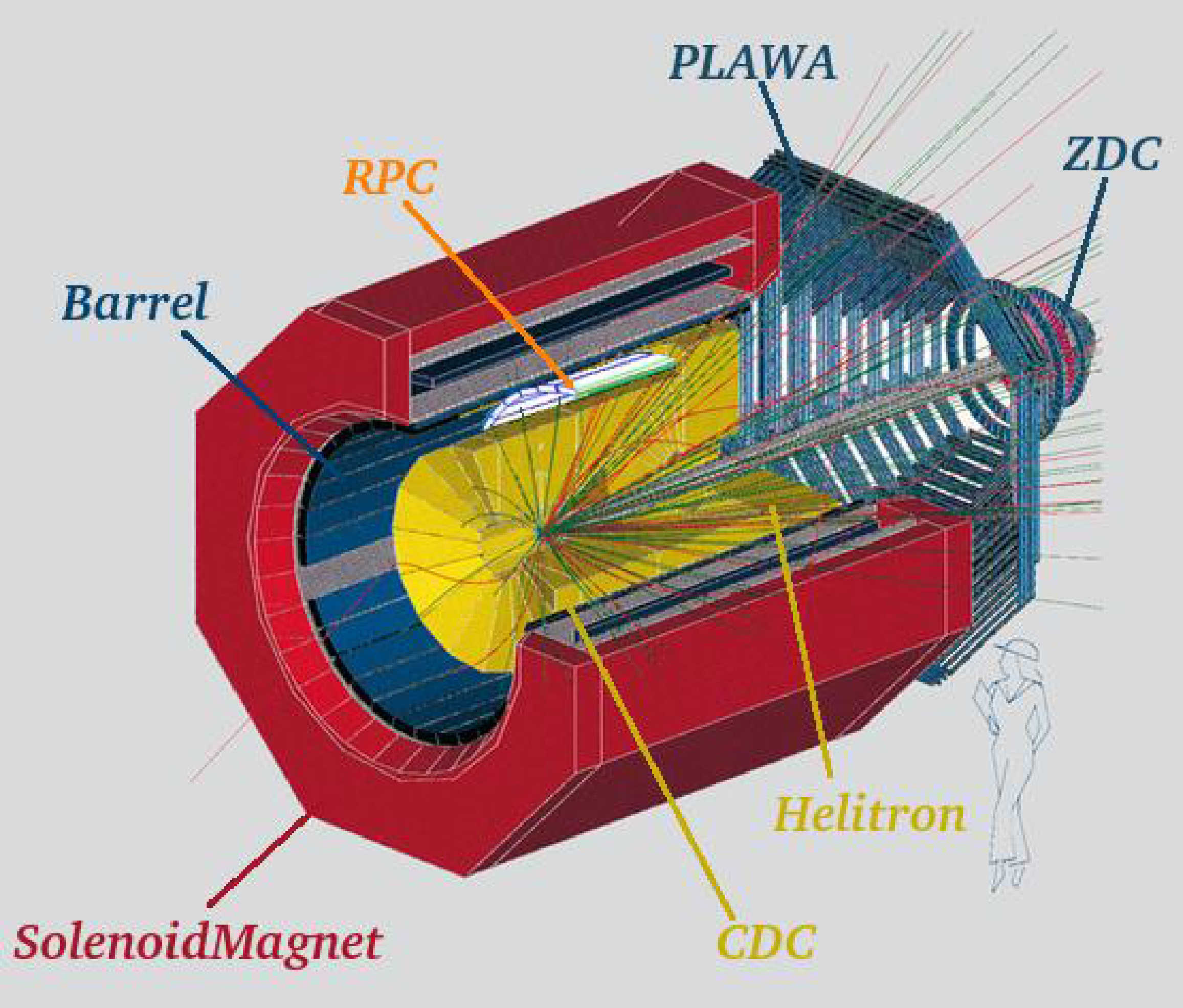}
\end{center}
\caption{The FOPI detector system: The drift chambers (yellow), the time of flight detectors Barrel,Plawa(blue) and RPC(white) surrounded by a solenoid magnet (red).}
\label{Fopi}
\end{figure}

In the last years there have been carried out an extensively discussion about the possible existence of kaonic nuclear states, especially the $K^{-}pp$. This basic kaonic nuclear state was first predicted by Akaishi and Yamazaki with a mass of M=2322 MeV/$c^{2}$, a binding energy of $B_{K}= 48 MeV$ and a width of $\Gamma= 61 MeV$ \cite{KNC1,KNC2}. The hypothesis has been put forward, that this stats is suggested to be populated quite favorably in a $p+p \rightarrow ppK^{-} + K^{+}$ reaction due to the large momentum transfer and the small impact parameter of the reaction. For this reason have planed an experiment to measure exclusively the suggested reaction $p+p\rightarrow ppK^{-} + K^{+} \rightarrow (\Lambda + p) + K^{+} \rightarrow \left( (\pi^{-}+p)+p\right)+ K^{+}$ with the FOPI detector system (Fig. \ref{Fopi}) at SIS18 at GSI (\cite{Prop}).
\section{The $\Lambda$-Trigger}
\subsection{Motivation}
For this exclusiv measurement of the $ppK^{-}$ an additional detector at FOPI is needed for the following purposes:
\begin{itemize}
\item{Enhancement of the events containing a $\Lambda$-Hyperon.}
\item{Improvement of the tracking for small polar angle in FOPI.}
\end{itemize}
\subsection{Assembly of the Trigger}
\begin{figure}[h]
\begin{center}
\includegraphics[width=6cm]{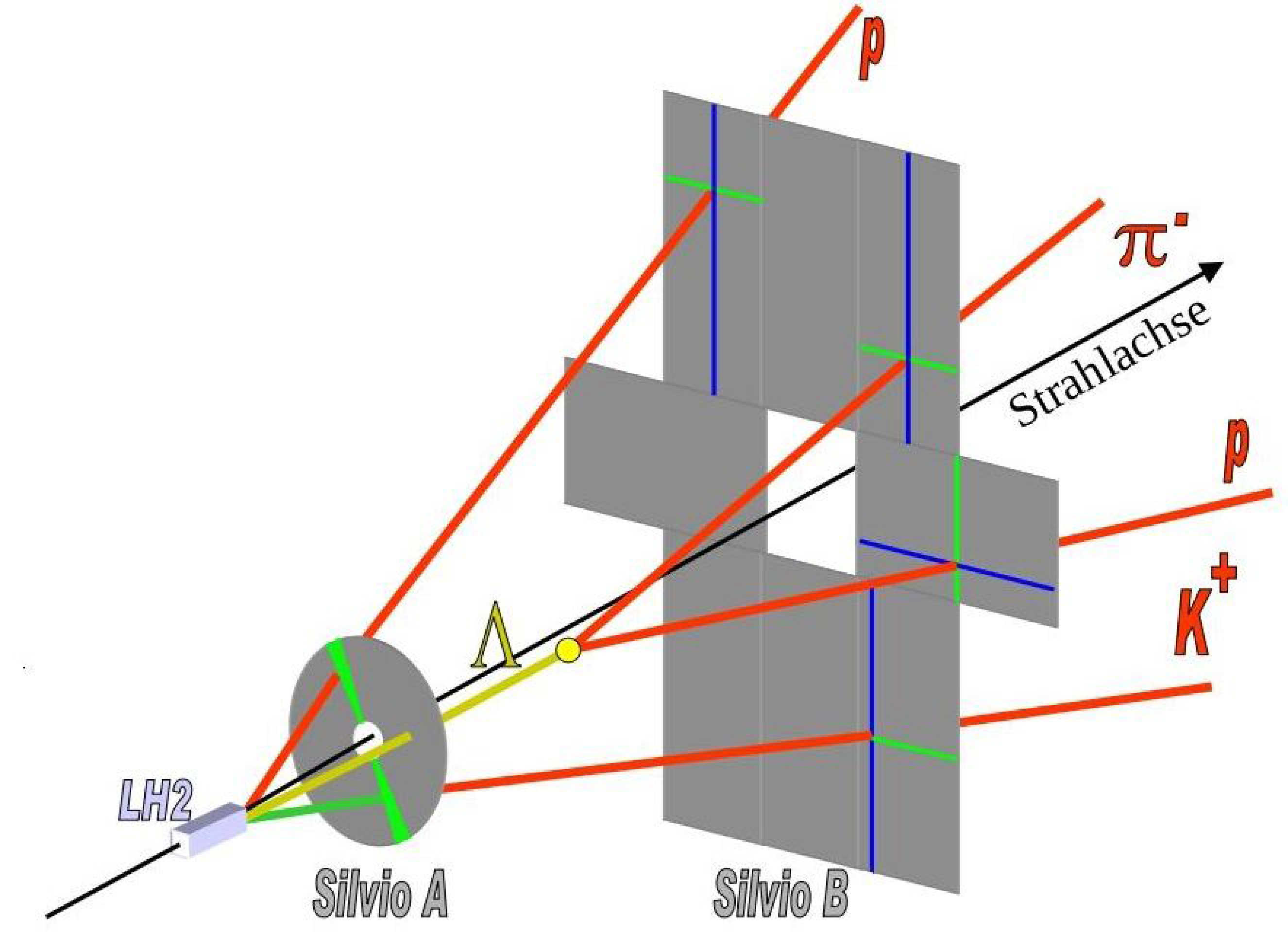}
\end{center}
\caption{Schematic picture of Si$\Lambda$ViO with a typical events with a $\Lambda$ decaying into p and $\pi^{-}$ between the two layers. Fired strips are indicated in blue and green.}
\label{Silviopic}
\end{figure}

These two requirements can be addressed with the detector Si$\Lambda$ViO (Silicon for $\Lambda$-Vertex and Identification Online). This detector contains two layers of 1 mm silicon detectors. Figure \ref{Silviopic} shows a schematic drawing of Si$\Lambda$ViO. The first layer (Si$\Lambda$ViO A) is an annular single-sided detector with 32 sectors. The second is a patchwork of 8 doubled-sided strip detectors with 16x60 strips each.
 The first layer is placed at a distance of 3 cm from the target and the second at 15.5 cm. Since the $\Lambda$-Hyperon has a decay length of 7.89 cm, more then 60\% of the $\Lambda$-Hyperons \cite{DPG}, produce in the target, will decay between these two layers. As more then 60\% of the $\Lambda$-Hyperon decay into $p+\pi^{-}$ \cite{DPG}, a higher hit multiplicity in Si$\Lambda$ViO B than in Si$\Lambda$ViO A will be recorded. Fig. \ref{Silviopic} shows such an events, indicating the different number of charged particle hits.\\
The readout of the 32 sectors of Si$\Lambda$Vio A and the 8$\times$16 strips of Si$\Lambda$ViO B, which are used for triggering, are read out by analog readout modules by the company Mesystec. These modules are able to generate a trigger signal depending on the number of channels with a signal higher than a set threshold. This trigger signal is generated about 150 ns after the corresponding particle has crossed the detector.\\
To improve the tracking in forward direction additional the position information delivered by Si$\Lambda$Vio B can be used. To achieve this, the 8 times 60 strips are read out by 4 APV multiplexer chips with a sampling rate of 40 MHz. This chip is able to store the signal for about 4 $\mu$s until a valid read-out trigger is received.
\section{Results of p+p-Test experiment}
\label{Results}
Two test experiments with proton beam were carried out at GSI to verify the performance of the Si$\Lambda$ViO system together with the whole FOPI detector system. During the first test the generation of the trigger signal was tested and its transport in the general FOPI acquisition. During the second test it was possible to accumulate enough statistics to test both the reconstruction capability of the $\Lambda$ hyperons and the trigger selection efficiency.  The final setup which includes a new start detector, a beam monitor and a veto detector, has been mounted and verified during the last test as well.
\begin{figure}[h]
\begin{center}
\includegraphics[width=5cm]{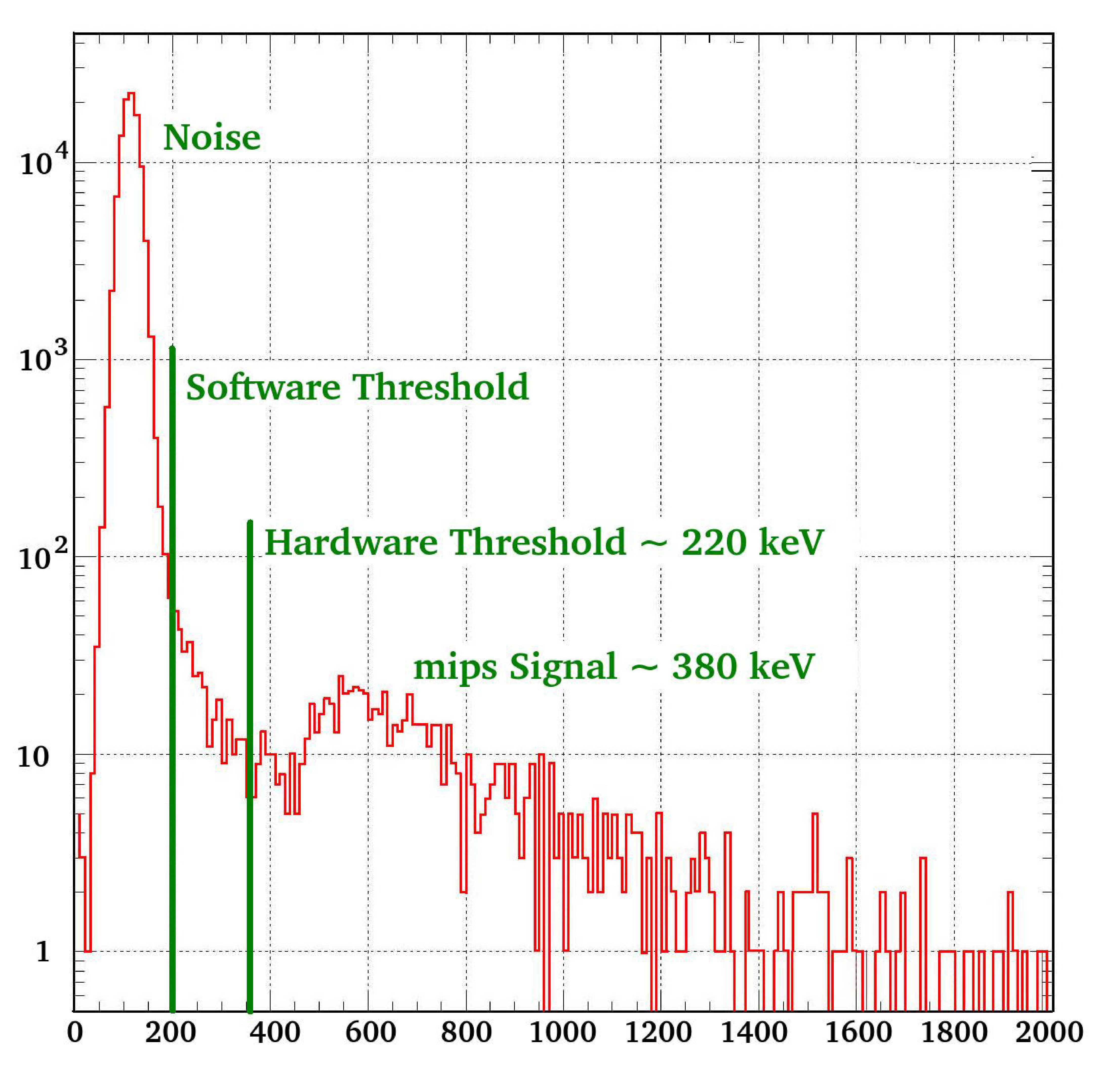}
\end{center}
\caption{Raw Spektrum of Analog-Mesystec-Readout. Green lines show the software threshold (mean+3.5$\sigma$ of noise) and the hardware threshold at about 220 keV.}
\label{Raw}
\end{figure}
\subsection{Signal Separation}
\label{Trigger}
The first thing that have to be clarified is, whether it is possible to separate the signal of mip from the noise.
Figure \ref{Raw} shows the raw spectrum of one strip of the analog readout of Si$\Lambda$ViO B. The green lines indicated the hardware threshold of about 220 keV of the trigger and the software threshold, with the value mean+3.5$\sigma$ of the noise.
\subsection{Rate Reduction}
\begin{figure}
\begin{center}
\includegraphics[width=5cm]{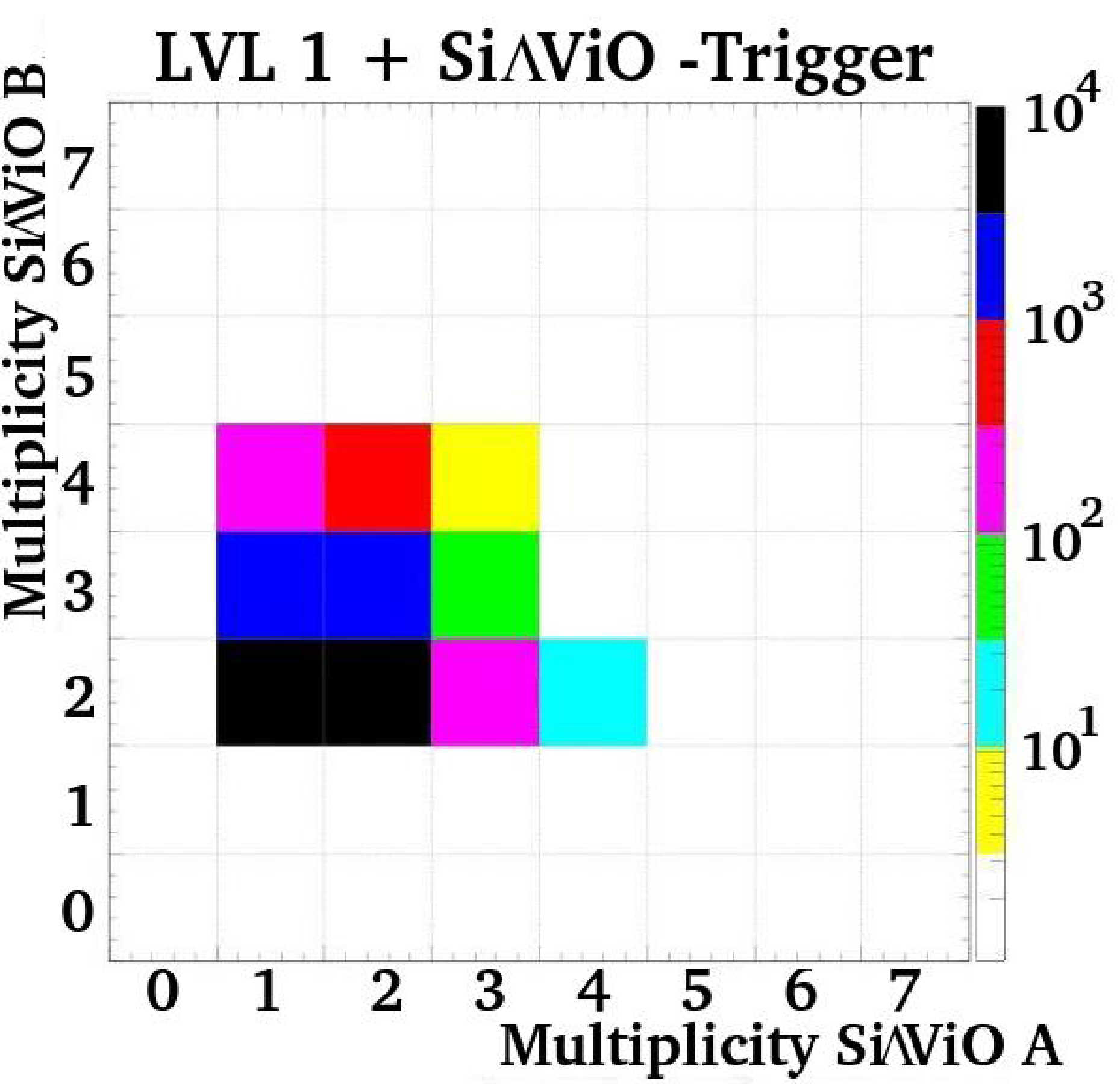}
\end{center}
\caption{Multiplicity of Si$\Lambda$ViO B versus Multiplicity of Si$\Lambda$ViO A for the LVL1 trigger of FOPI + Si$\Lambda$ViO Trigger.}
\label{CM}
\end{figure}

One of the main goal of the experiments was to find out the rate reduction capability of Si$\Lambda$ViO. To do this we compared the trigger rate of the LVL1-Trigger, which requires at least two and at most 5 charged particles in the time of flight detectors, with the LVL1-Trigger+Si$\Lambda$ViO, which requires one or two particles in the Si$\Lambda$ViO A layer and two till four in Si$\Lambda$ViO B. With the final setup we achieved a rate reduction of about 27. To ensure that Si$\Lambda$ViO is cutting on the correct multiplicity windows we looked on the multiplicity correlation in Silvio B versus Silvio A (Figure \ref{CM}). From these correlation it can be seen, that the signal is reduced quite good  to the expected hardware multiplicity windows.\\
We also compared the reduction we achieved with a target and without a target. By this we find out, that less then 13\% of the events, triggered by Si$\Lambda$ViO, are not coming from the target \cite{Maddin}.
\section{Summary and Outlook}
\label{Summary}
In the test experiments we have been able to show, that Si$\Lambda$ViO is able to trigger on the signal of mip particles. Furthermore we have seen, that it is possible to reduce the LVL1 rate with Si$\Lambda$ViO of about a factor 29, while more then 97\% are events coming from the target\\
In the near future we will find out, whether we are able to improve the tracking to reconstruct $\Lambda$-Hyperonen in low polar angles.\\
The final production beam time will be in August 2009.



\end{document}